\documentclass[aps,prd,preprint,nofootinbib,11pt]{revtex4}
\usepackage{amsfonts}
\usepackage{mathrsfs}
\usepackage{graphicx}
\usepackage{amsmath}
\usepackage{amssymb}
\usepackage{subfigure}
\usepackage{epsfig}
\usepackage{graphicx}
\usepackage{ulem}
\usepackage{color}
\usepackage{soul}
\newcommand{\bqa}{\begin{eqnarray}}
\newcommand{\eqa}{\end{eqnarray}}
\newcommand{\beq}{\begin{equation}}
\newcommand{\eeq}{\end{equation}}
\allowdisplaybreaks[1]
\graphicspath{{fig/}{dia/}} \DeclareGraphicsExtensions{.eps}

\begin{document}
\title{\Large Possible structure of the newly found exotic state $\eta_1(1855)$\\[7mm]}

\author{Bing-Dong Wan$^{1,2}$\footnote{wanbingdong16@mails.ucas.ac.cn}, Sheng-Qi Zhang$^2$\footnote{zhangshengqi20@mails.ucas.ac.cn} and Cong-Feng Qiao$^{2}$\footnote{qiaocf@ucas.ac.cn, corresponding author}\vspace{+3pt}}

\affiliation{$^1$  School of Fundamental Physics and Mathematical Sciences, Hangzhou Institute for Advanced Study, UCAS, Hangzhou 310024, China\\
$^2$ School of Physics, University of Chinese Academy of Science, Yuquan Road 19A, Beijing 100049, China}

\author{~\\~\\}

\begin{abstract}
\vspace{0.3cm}
Recently, a hadronic state named $\eta_1(1855)$, about 1.86 GeV, was observed in the BESIII experiment. This finding has a peculiar interest due to its exotic quantum number $J^{PC}=1^{-+}$. In this paper, we examine the tetraquark interpretation for the structure of $\eta_1(1855)$ in the configurations of $[1_c]_{\bar{s}s}\otimes[1_c]_{\bar{q}q}$ and $[1_c]_{\bar{s}q}\otimes[1_c]_{\bar{s}q}$, and perform a mass spectrum calculation in the framework of QCD sum rules. The results show that the observed $\eta_1(1855)$ could be embedded into the $[1_c]_{\bar{s}s}\otimes[1_c]_{\bar{q}q}$ configuration. The possible tetraquark and hybrid decay modes are analyzed, which are critical in decoding its inner structure. In the same way, we evaluate as well the $\bar{s}s\bar{s}s$ molecular state with $1^{-+}$ quantum number and find that there might exist two such ground states.
 \end{abstract}
 \maketitle
\newpage

\section{Introduction}

Investigation on hadronic states beyond the conventional quark model \cite{GellMann:1964nj,Zweig}, such as multiquarks, hybrids, and glueballs, may greatly enrich the hadron family and our knowledge of the nature of quantum chromodynamics(QCD). So far, more than thirty such states or candidates have been observed in experiment ever since the observation of X(3872) \cite{Choi:2003ue}. It is highly expected that a great more new hadronic states will emerge soon afterwards, implying the renaissance of hadron physics. To decoding the hadronic structure of the new experimental observations is one of the intriguing and important topics in hadron physics.

In the light hadron sector, partly due to the small spacings between various states and hence large mixing, it is usually hard to discriminate new hadronic states (like multiquarks, hybrids, glueballs and so on) from conventional ones in experiments, except the former possess some exotic quantum numbers (like $0^{--}$, $0^{+-}$, $\cdots$). With the accumulation of experimental data in charm quark region, BESIII is ready to examine carefully the hadron physics in this area \cite{BESIII:2010gmv,BES:2003aic,BES:2005ega,BESIII:2010vwa,BESIII:2019wkp,BESIII:2016qzq,BESIII:2020vtu, BESIII:2017kqw,BESIII:2017hyw,BESIII:2019cuv}, including of course exotic hadrons.

Very recently, by analyzing the partial wave of the process $J/\psi\to\gamma\eta\eta^\prime$, the BESIII Collaboration observed a structure at $1.855$ GeV, named $\eta_1(1855)$, in the $\eta$ and $\eta^\prime$ invariant mass spectrum with $19\;\sigma$ significance and its decay width is some $188$ MeV \cite{BESIII:2022riz,BESIII:2022qzu}. This finding soon attracts a lot of attention, since $\eta_1(1855)$ possesses an exotic quantum number of $J^{PC}=1^{-+}$. In the past there were couple of exotic-hadron candidates observed, while no one is so well determined as this guy. To understand more of the $\eta_1(1855)$'s property is a tempting and pressing problem in hadron physics.

In the literature, the hybrid explanation of $\eta_1(1855)$ was proposed by flux tube model~\cite{Qiu:2022ktc}, QCD sum rules~\cite{Chen:2022qpd} and effective Lagrangian method~\cite{Shastry:2022mhk}. Hybrid, carrying valence quarks and gluonic degrees of freedom, can possess exotic quantum number which cannot be reached by a $ q\bar{q} $ state. The lowest-lying hybrids are expected to have $ J^{PC}=1^{-+} $, which have been explored in a variety of theories~\cite{Meyer:2015eta, Klempt:2007cp}. The lattice calculations of the $ 1^{-+} $ hybrid were made in the quenched approximation, where the results predicted that the $ 1^{-+} $ nonet of hybrids were the lightest with a mass from 1.8 GeV to 2.1 GeV~\cite{Harter:1997jq, Lacock:1998be, Mei:2002ip, Hedditch:2005zf}. Within the MIT bag model, the mass of $ 1^{-+} $ hybrids were about $ 1.8 \pm 0.4 $ GeV and the hybrid phenomenology was also carried out in Ref.~\cite{Chanowitz:1982qj}. Constituent gluon models,  assuming a diagonal gluon angular momentum, predicted that the mass of light exotic hybrids were 1.8-2.2 GeV~\cite{Ishida:1991mx}, which was consistent with the lattice results. In the framework of QCD sum rules, the estimation for the $ 1^{-+} $ hybrid was about 1.6-1.8 GeV~\cite{Latorre:1985tg, Chetyrkin:2000tj, Narison:2009vj}.

However, in physics, the exotic quantum number $J^{PC}=1^{-+}$ can also be assigned to a tetraquark state. In fact, the dynamical gluon of the hybrid can easily split into a quark pair 
so that to discriminate the different explanations of $ \eta_1(1855)$ as hybrid or pure tetraquark one needs to know more fine structures of the observed state. About $1^{-+}$ tetraquark there have been some investigations in the literature \cite{Chen:2008ne,Narison:2009vj,Dong:2022cuw,Yang:2022lwq}.
 In Ref. \cite{Chen:2008ne}, the $1^{-+}$ isoscalar tetraquark states with diquark-antidiquark configurations in light sector were studied in QCD sum rules by Chen {\it et al.}, and the masses were found around $1.8$-$2.1$ GeV.
 Narison investigated  the $1^{-+}$ tetraquarks in diquark-antidiquark and molecule configurations but with no strange quark component, and the numerical results are around $1.7$ GeV~\cite{Narison:2009vj}. In Refs \cite{Dong:2022cuw,Yang:2022lwq}, $\eta_1(1855)$ was interpreted as a $K\bar{K}_1(1400)$ molecular state by using the one meson exchange model.

In this work, the masses of the light isoscalar tetraquark states in configurations of $[1_c]_{\bar{s}s}\otimes [1_c]_{\bar{q}q}$ and $[1_c]_{\bar{s}q}\otimes [1_c]_{\bar{s}q}$ with $J^{PC}=1^{-+}$ are investigated by means of Shifman, Vainshtein and Zakharov (SVZ) sum rules technique \cite{Shifman}. The SVZ sum rules, viz the QCD sum rules (QCDSR), has some peculiar advantages in exploring hadron properties involving nonperturbative QCD. QCDSR is a QCD based theoretical framework which incorporates nonperturbative effects and has already achieved a lot in the study of hadron spectroscopy\cite{Xin:2021wcr, Wang:2021qus, Qiao:2013dda, Qiao:2013raa, Tang:2016pcf,Tang:2019nwv, Wan:2019ake, Wan:2020fsk,Wan:2020oxt, Wan:2021vny, Albuquerque:2013ija,Matheus:2006xi,Wang:2013vex,Cui:2011fj,Narison:2002pw,P.Col,Reinders:1984sr}. The starting point to establish the sum rules is to construct the proper interpolating currents corresponding to hadrons of interest, which possesses the foremost information of the concerned hadrons, like quantum numbers and structure components. The two-point correlation function is constructed by interpolating currents, it has two representations: the operator product expansion (OPE) representation and the phenomenological representation. By equating these two representations, the QCD sum rules will be formally established, from which the hadron mass may be deduced.

The rest of the paper is arranged as follows. After the Introduction, a brief interpretation of QCD sum rules and some primary formulas in our calculation are presented in Sec. \ref{Formalism}. The numerical analysis and the possible tetraquark and hybrids decay modes are given in Sec. \ref{Numerical}. The last part is left for a brief summary.

\section{Formalism}\label{Formalism}

The lowest order currents for light tetraquark states with $J^{PC}=1^{-+} $ in molecular configuration are found to be in forms:
\begin{eqnarray}\label{current_lambda}
j_\mu^A(x)&=&i [\bar{s}_a(x)\gamma_5 s_a(x)][\bar{q}_b(x) \gamma^\mu \gamma_5 q_b(x)] \;,\label{Ja}\\
j_\mu^B(x)&=&i [\bar{s}_a(x)\gamma^\mu \gamma_5 s_a(x)][\bar{q}_b(x) \gamma_5 q_b(x)] \;,\label{Jb}\\
j_\mu^C(x)&=& \frac{i}{\sqrt{2}} {\Big\{}[\bar{s}_a(x)\gamma_5 q_a(x)][\bar{q}_b(x) \gamma^\mu \gamma_5 s_b(x)] \nonumber\\
&+& [\bar{s}_a(x)\gamma^\mu\gamma_5 q_a(x)][\bar{q}_b(x)  \gamma_5 s_b(x)] {\Big\}}\;,\label{Jc}\\
j_\mu^D(x)&=&\frac{1}{\sqrt{2}} {\Big\{}[\bar{s}_a(x) q_a(x)][\bar{q}_b(x) \gamma^\mu  s_b(x)] \nonumber\\
&-& [\bar{s}_a(x)\gamma^\mu q_a(x)][\bar{q}_b(x)   s_b(x)] {\Big\}}\;, \label{Jd}
\end{eqnarray}
where the subscripts $a$ and $b$ are color indices, $q$ stands for light quarks. In our calculation, $q(x)=1/\sqrt{2}[u(x)+d(x)]$ for isoscalar state.
Hereafter, for simplicity the four $1^{-+}$ currents in Eqs. (\ref{Ja})–(\ref{Jd}) will be referred as cases $A$ to $D$, respectively.

With the currents (\ref{Ja})-(\ref{Jd}), the two-point correlation function can be readily established, i.e.,
\begin{eqnarray}
\Pi_{\mu\nu}(q^2) &=& i \int d^4 x e^{i q \cdot x} \langle 0 | T \{ j_\mu (x),\;  j_\nu^\dagger (0) \} |0 \rangle \;,
\end{eqnarray}
where $ |0 \rangle$ denotes the physical vacuum. The correlation function has the following Lorentz covariance form:
\begin{eqnarray}
\Pi_{\mu\nu}(q^2) &=&-\Big( g_{\mu \nu} - \frac{q_\mu q_\nu}{q^2}\Big) \Pi_1(q^2)+ \frac{q_\mu q_\nu}{q^2}\Pi_0(q^2)\;,
\end{eqnarray}
where the subscripts $1$ and $0$, respectively, denote the quantum numbers of the spin 1 and 0 mesons.

On the phenomenological side, the correlation function $\Pi(q^2)$ can be expressed as a dispersion integral over the physical regime after isolating the ground state contribution from the tetraquark state, i.e.,
 \begin{eqnarray}
\Pi^{phen}_i(q^2) & = & \frac{\lambda_i^2}{M_i^2 - q^2} + \frac{1}{\pi} \int_{s_0}^\infty d s \frac{\rho_i(s)}{s - q^2} \; , \label{hadron}
\end{eqnarray}
where the subscripts $i$ runs from $A$ to $D$, $M$ denotes the tetraquark mass, $\lambda$ is the coupling constant of current to hadron, and $\rho(s)$ is the spectral density that contains the contributions from higher excited states and the continuum states above the threshold $s_0$.

In the OPE representation, the dispersion relation can express the correlation function $\Pi(q^2)$ as
 \begin{eqnarray}
\Pi^{OPE}_{i} (q^2) = \int_{s_{min}}^{\infty} d s
\frac{\rho^{OPE}_{i}(s)}{s - q^2}+\Pi_i^{sum}(q^2)\; .
\label{OPE-hadron}
\end{eqnarray}
Here, $s_{min}$ is the kinematic limit, which usually corresponds to the square of the sum of current-quark masses of the hadron \cite{Albuquerque:2013ija}, $\Pi^{sum}_i$ is the sum of those contributions in correlation function that have no imaginary part but are nontrivial after the Borel transformation, $\rho^{OPE}_{i}(s) = \text{Im} [\Pi^{OPE}_{i}(s)] / \pi$, and
\begin{eqnarray}
\rho^{OPE}(s) & = & \rho^{pert}(s) + \rho^{\langle \bar{q} q
\rangle}(s) +\rho^{\langle G^2 \rangle}(s) + \rho^{\langle \bar{q} G q \rangle}(s)\nonumber\\
&+& \rho^{\langle \bar{q} q \rangle^2}(s)
+ \rho^{\langle G^2 \rangle\langle \bar{q} q \rangle}(s)
+ \rho^{\langle \bar{q} q \rangle\langle \bar{q} G q \rangle}(s) \;. \label{rho-OPE}
\end{eqnarray}

To calculate the spectral density of the OPE side, Eq. (\ref{rho-OPE}), the light quark full propagators $S_{ij}^q(x)$ is employed, say
\begin{eqnarray}
S^q_{i j}(x) \! \! & = & \! \! \frac{i \delta_{i j} x\!\!\!\slash}{2 \pi^2
x^4} - \frac{\delta_{i j} m_q}{4 \pi^2 x^2} - \frac{i t^a_{i j} G^a_{\alpha\beta}}{2^{5}\; \pi^2 x^2}(\sigma^{\alpha \beta} x\!\!\!\slash
+ x\!\!\!\slash \sigma^{\alpha \beta}) - \frac{\delta_{i j}}{12} \langle\bar{q} q \rangle + \frac{i\delta_{i j}
x\!\!\!\slash}{48} m_q \langle \bar{q}q \rangle - \frac{\delta_{i j} x^2}{192} \langle g_s \bar{q} \sigma \cdot G q \rangle \nonumber \\ &+& \frac{i \delta_{i j} x^2 x\!\!\!\slash}{2^7\times 3^2\;} m_q \langle g_s \bar{q} \sigma \cdot G q \rangle - \frac{t^a_{i j} \sigma_{\alpha \beta}}{192}
\langle g_s \bar{q} \sigma \cdot G q \rangle
+ \frac{i t^a_{i j}}{768} (\sigma_{\alpha \beta} x \!\!\!\slash + x\!\!\!\slash \sigma_{\alpha \beta}) m_q \langle
g_s \bar{q} \sigma \cdot G q \rangle \;,
\end{eqnarray}
where the vacuum condensates are clearly displayed. For more explanation on above propagator, readers may refer to Refs.~\cite{Wang:2013vex, Albuquerque:2013ija}. The Feynman diagrams corresponding to each term of Eq. (\ref{rho-OPE}) are schematically shown in Fig. \ref{feyndiag}, and the analytical formulas are given in the appendix.

\begin{figure}
\includegraphics[width=8.8cm]{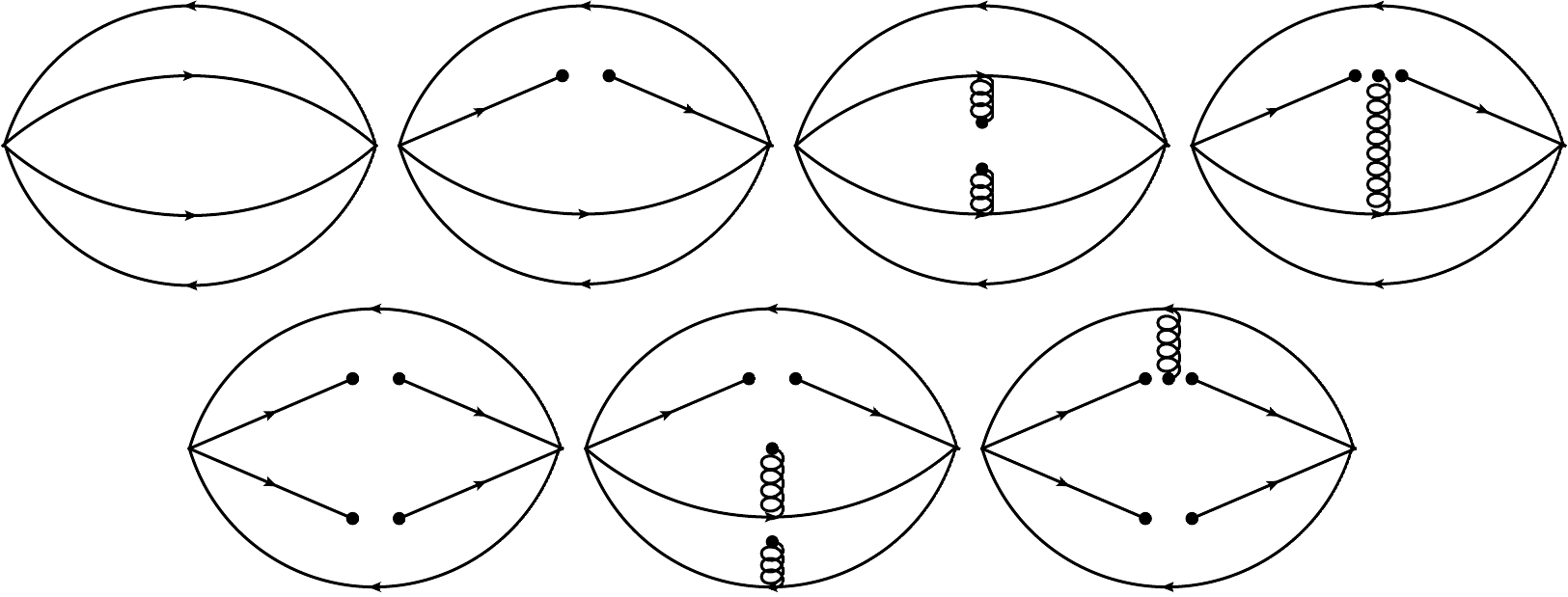}
\caption{The typical Feynman diagrams related to the correlation function, where the solid lines stand for the quarks and the spiral ones for gluons.} \label{feyndiag}
\end{figure}

Performing the Borel transform on Eqs. (\ref{hadron}) and (\ref{OPE-hadron}), and matching the OPE side with the phenomenological side of the correlation function $\Pi(q^2)$, one can finally obtain the mass of the tetraquark state,
\begin{eqnarray}
M(s_0, M_B^2) = \sqrt{- \frac{L_{1}(s_0, M_B^2)}{L_{0}(s_0, M_B^2)}} \; . \label{mass-Eq}
\end{eqnarray}
Here $L_0$ and $L_1$ are respectively defined as
\begin{eqnarray}
L_{0}(s_0, M_B^2) =  \int_{s_{min}}^{s_0} d s \; \rho^{OPE}(s) e^{-
s / M_B^2} +\Pi^{sum}(M_B^2)  \;,  \label{L0}
\end{eqnarray}
and
\begin{eqnarray}
L_{1}(s_0, M_B^2) =
\frac{\partial}{\partial{\frac{1}{M_B^2}}}{L_{0}(s_0, M_B^2)} \; .
\end{eqnarray}

\section{Numerical analysis}\label{Numerical}
\subsection{Mass spectrum calculation}
In performing the numerical calculation, the broadly accepted inputs are taken \cite{pdg,Matheus:2006xi, Cui:2011fj, Narison:2002pw,P.Col,Tang:2019nwv}, i.e., $m_u=2.16^{+0.49}_{-0.26}\; \text{MeV}$, $m_d=4.67^{+0.48}_{-0.17}\; \text{MeV}$, $m_s=(95\pm5)\; \text{MeV}$, $\langle \bar{q} q \rangle = - (0.23 \pm 0.03)^3 \; \text{GeV}^3$, $\langle \bar{s} s \rangle=(0.8\pm0.1)\langle \bar{q} q \rangle$, $\langle \bar{q} g_s \sigma \cdot G q \rangle = m_0^2 \langle\bar{q} q \rangle$, $\langle \bar{s} g_s \sigma \cdot G s \rangle = m_0^2 \langle\bar{s} s \rangle$, $\langle g_s^2 G^2 \rangle = (0.88\pm0.25) \; \text{GeV}^4$, and $m_0^2 = (0.8 \pm 0.2) \; \text{GeV}^2$.

Moreover, there exist two additional parameters $s_0$ and $M_B^2$ introduced in establishing the sum rules, which can be fixed in light of the so-called standard procedures by fulfilling the following two criteria \cite{Shifman,Reinders:1984sr, P.Col,Albuquerque:2013ija}. The first one asks for the convergence of the OPE, which is to compare relative contribution of higher dimension condensate to the total contribution on the OPE side, and then a reliable region for $M_B^2$ will be chosen to retain the convergence. The other criterion of QCD sum rules requires that the pole contribution (PC) is more than $(40-60)\%$ of the total \cite{Xin:2021wcr,Wang:2021qus}. Mathematically, the two criteria can be formulated as:
\begin{eqnarray}
  R^{OPE}= \left| \frac{L_{0}^{dim=8}(s_0, M_B^2)}{L_{0}(s_0, M_B^2)}\right|\, ,
\end{eqnarray}
\begin{eqnarray}
  R^{PC} = \frac{L_{0}(s_0, M_B^2)}{L_{0}(\infty, M_B^2)} \; . \label{RatioPC}
\end{eqnarray}

In order to find a proper value for continuum threshold $s_0$, a similar analysis as in Refs. \cite{Reinders:1984sr, P.Col,Albuquerque:2013ija} is performed. Therein, one needs to find the proper value, which has an optimal window for the mass curve of the light tetraquark state. Within this window, the physical quantity, that is the mass of the light tetraquark state, should be independent of the Borel parameter $M_B^2$ as much as possible. In practice, we will vary $\sqrt{s_0}$ by $0.1$ GeV to obtain the lower and upper bounds, and hence the uncertainties of $\sqrt{s_0}$ \cite{Albuquerque:2013ija}.

\begin{figure}
\includegraphics[width=6.8cm]{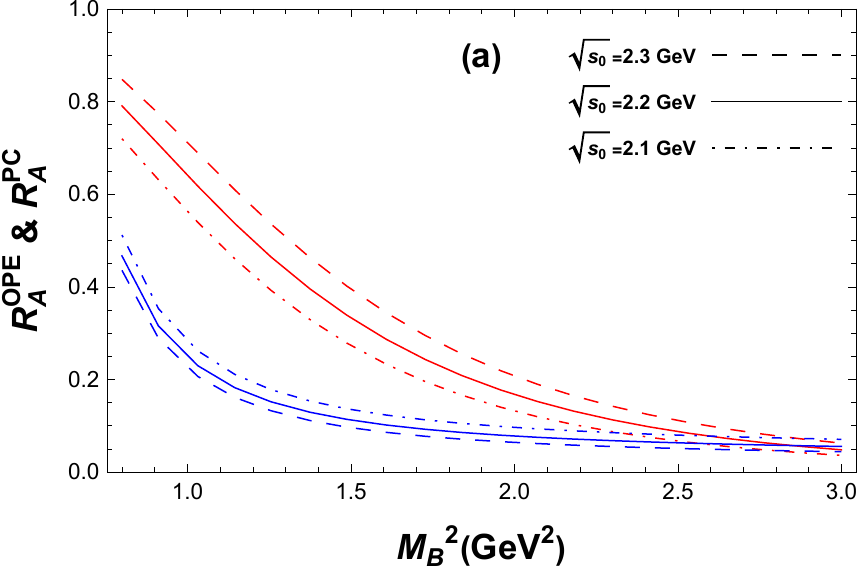}
\includegraphics[width=6.8cm]{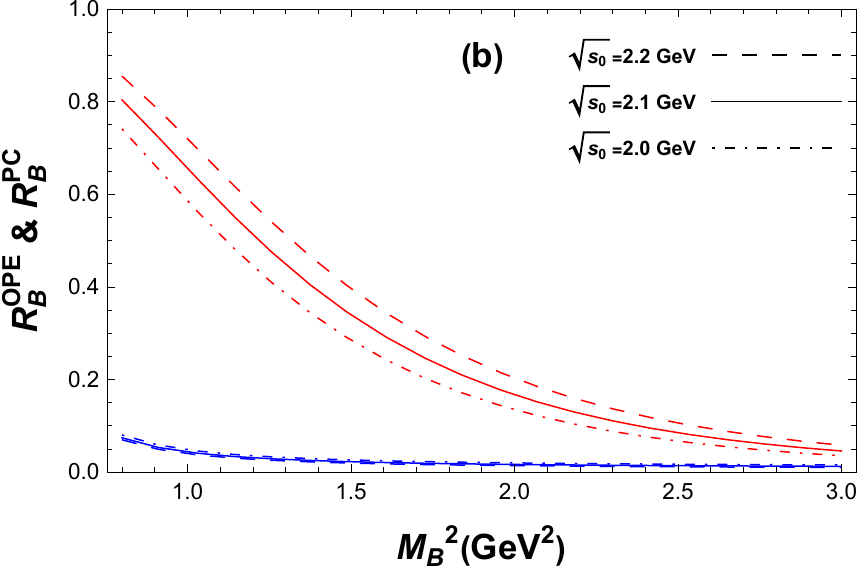}
\includegraphics[width=6.8cm]{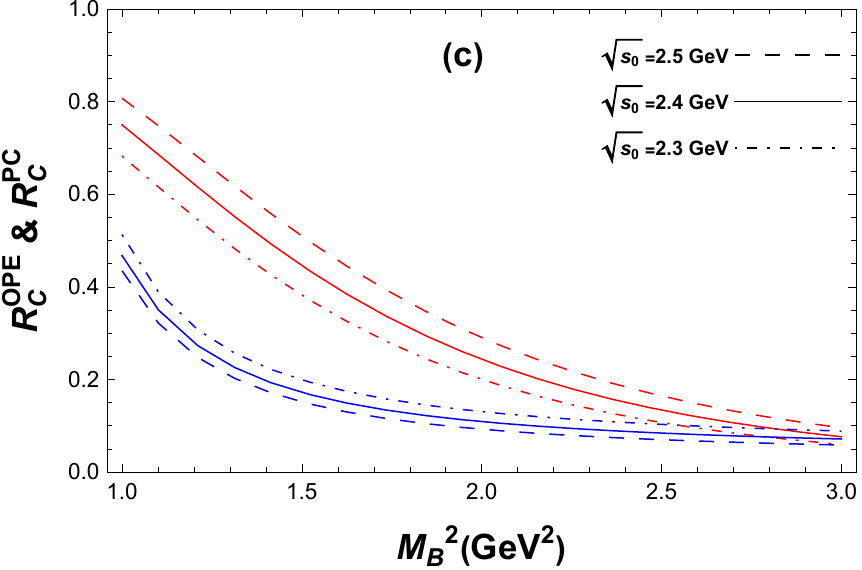}
\includegraphics[width=6.8cm]{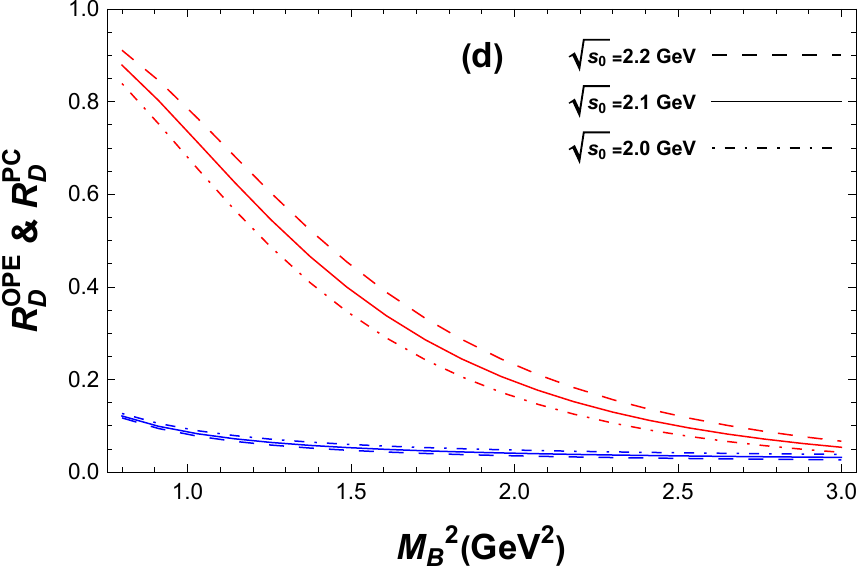}
\caption{ The ratios of ${R_{X}^{OPE}}$ and ${R_{X}^{PC}}$ as functions of the Borel parameter $M_B^2$ for different values of $\sqrt{s_0}$, where blue lines represent ${R_{X}^{OPE}}$ and red lines denote ${R_{X}^{PC}}$. Here, the subscripts $X$ runs from $A$ to $D$. (a), (b), (c), and (d) are for the current in Eqs.~(\ref{Ja}), (\ref{Jb}), (\ref{Jc}), and (\ref{Jd}), respectively.} \label{figA}
\end{figure}

\begin{figure}
\includegraphics[width=6.8cm]{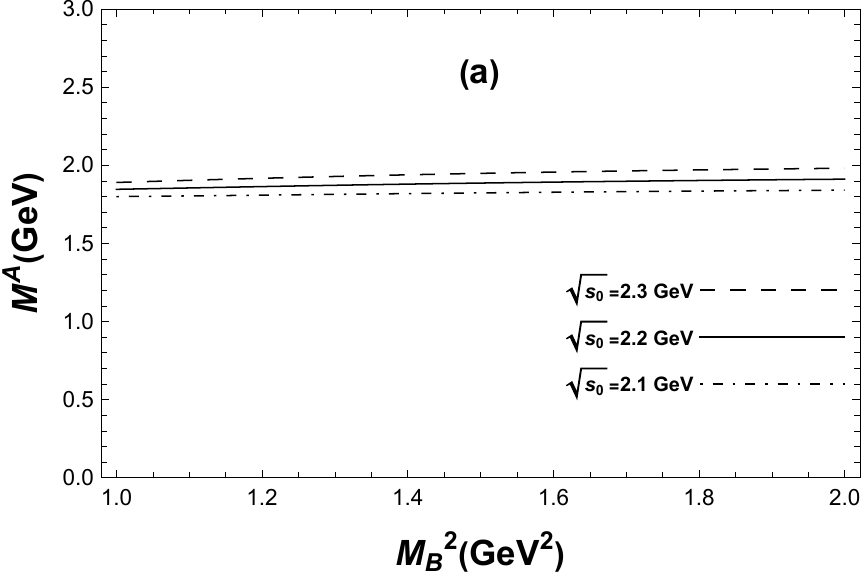}
\includegraphics[width=6.8cm]{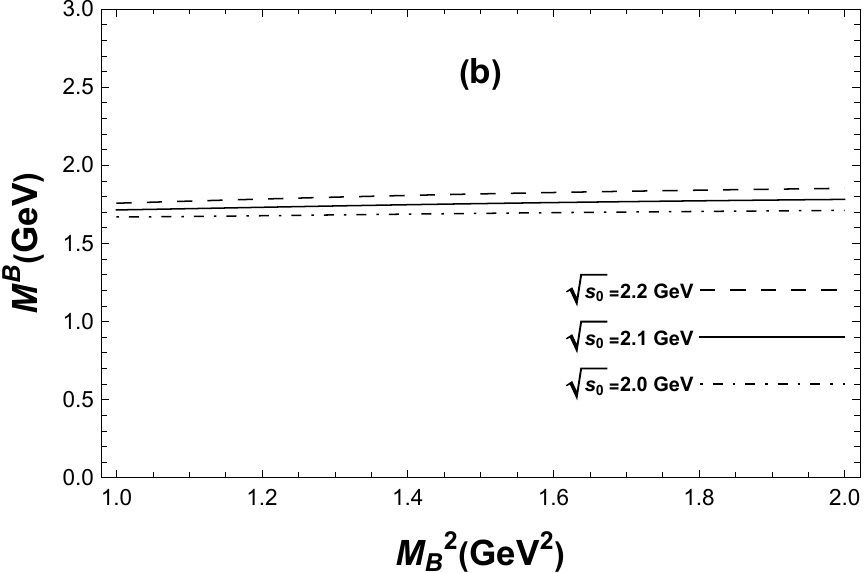}
\includegraphics[width=6.8cm]{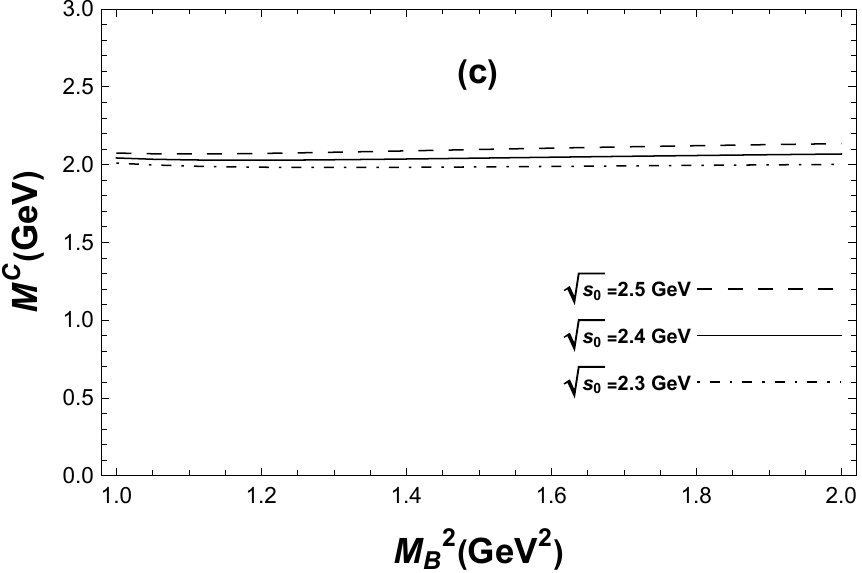}
\includegraphics[width=6.8cm]{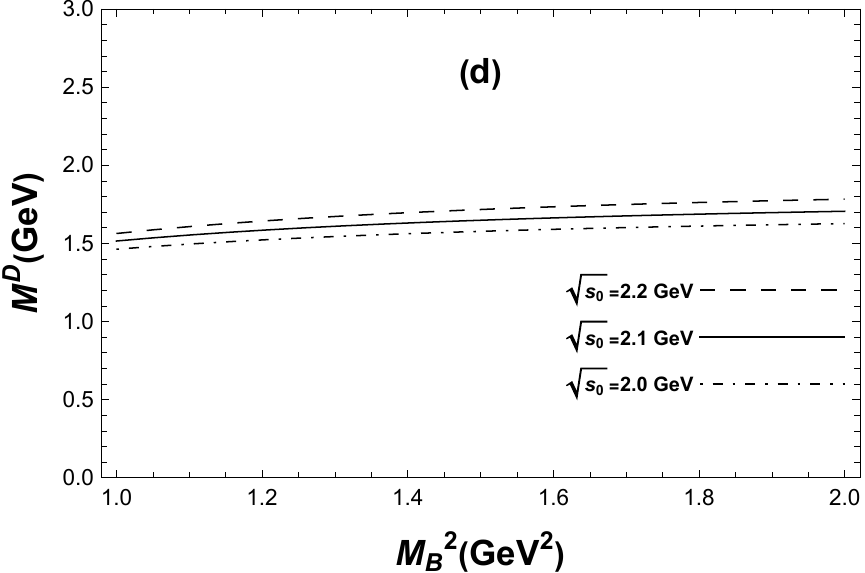}
\caption{The mass $M^{X}$ as a function of the Borel parameter $M_B^2$ for different values of $\sqrt{s_0}$. Here, the subscripts $X$ runs from $A$ to $D$. (a), (b), (c), and (d) are for the current in Eqs.~(\ref{Ja}), (\ref{Jb}), (\ref{Jc}), and (\ref{Jd}), respectively.} \label{figB}
\end{figure}

With the above preparation the mass spectrum of light tetraquaek states can be numerically evaluated. As an example, the ratios $R^{OPE}_{A}$ and $R^{PC}_{A}$ are presented as functions of Borel parameter $M_B^2$ in Fig. \ref{figA}(a) with different values of $\sqrt{s_0}$, i.e., $2.1$, $2.2$ and $2.3$ GeV. The reliant relations of $M^{A}$ on parameter $M_B^2$ are displayed in Fig. \ref{figB}(a). The optimal Borel window is found in range $1.1 \le M_B^2 \le 1.6\; \text{GeV}^2$, and the mass $M^{A}$ can then be obtained:
\begin{eqnarray}
M^{A} &=& (1.87\pm 0.08)\; \text{GeV}.\label{m1}
\end{eqnarray}
With the same analyses, and the OPE, pole contribution and the masses as functions of Borel parameter $M_B^2$ can be found in Fig.~\ref{figA} and \ref{figB}, respectively, the masses $M^{B}$, $M^{C}$ and $M^{D}$ can be extracted as follows:
\begin{eqnarray}
M^{B} &=& (1.75\pm 0.08)\; \text{GeV},\;\label{m2}\\
M^{C} &=& (2.05\pm 0.07)\; \text{GeV},\;\label{m3}\\
M^{D} &=& (1.63\pm 0.12)\; \text{GeV}.\;\label{m4}
\end{eqnarray}

\begin{figure}
\includegraphics[width=6.8cm]{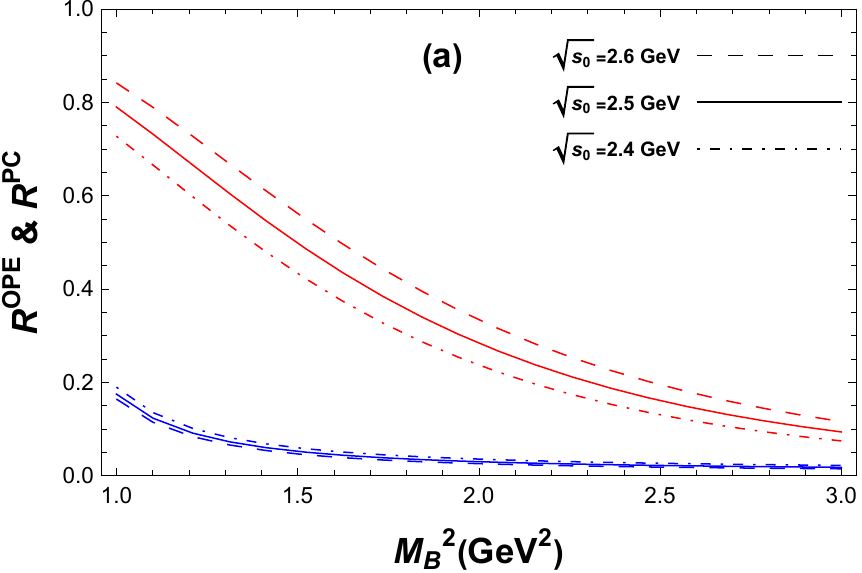}
\includegraphics[width=6.8cm]{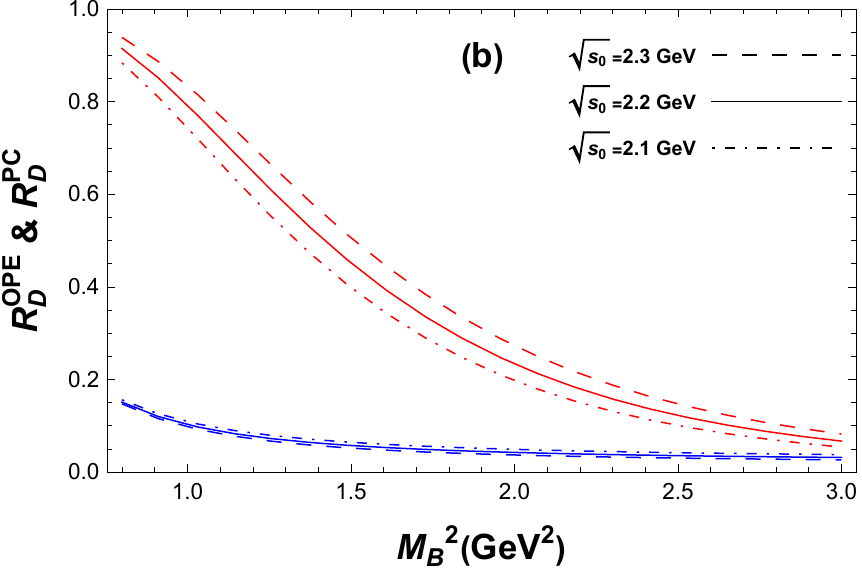}
\caption{ The same caption as in Fig \ref{figA}, but for the tetrastrange states. (a) and (b) are for the current in Eqs.~(\ref{Ja}) and (\ref{Jd}), respectively.} \label{figsA}
\end{figure}

\begin{figure}
\includegraphics[width=6.8cm]{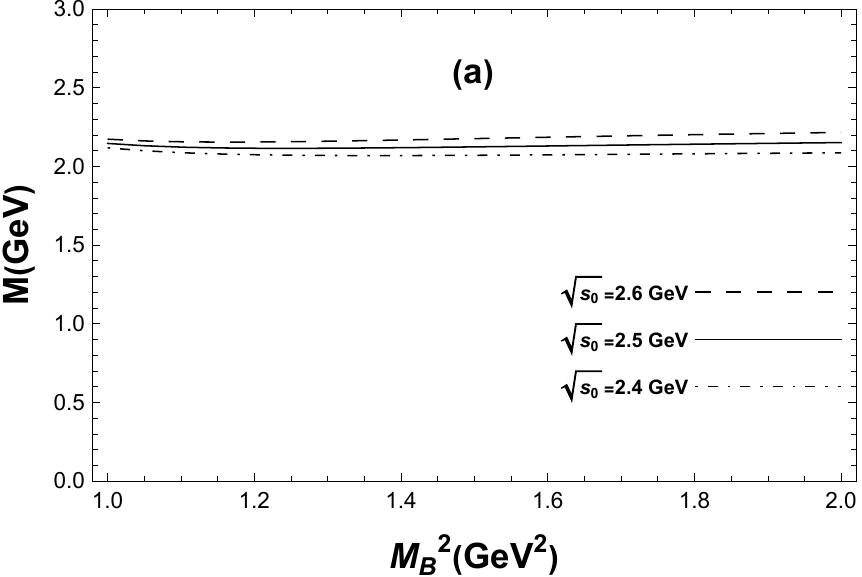}
\includegraphics[width=6.8cm]{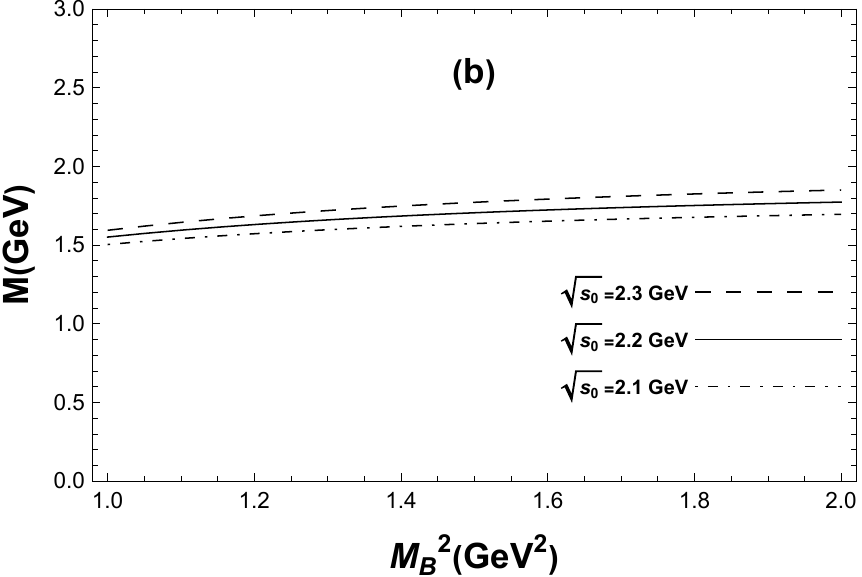}
\caption{ The same caption as in Fig \ref{figB}, but for the tetrastrange states. (a) and (b) are for the current in Eqs.~(\ref{Ja}) and (\ref{Jd}), respectively.} \label{figsD}
\end{figure}

Similarly, we can evaluate the $1^{-+}$ strange tetraquark states in configuration $[1_c]_{\bar{s}s}\otimes[1_c]_{\bar{s}s}$. With the replacement of light quark by strange quark in the obtained analytical results, the corresponding masses of strange tetraquark are readily obtained, that is
\begin{eqnarray}
M^A_{4s} &=& (2.14\pm 0.07)\; \text{GeV}\;,\\\label{m5}
M^D_{4s} &=& (1.71\pm 0.11)\; \text{GeV}\;,\label{m6}
\end{eqnarray}
The convergence of the OPE, pole contribution and the masses of tetrastrange states are shown as functions of the Borel parameter $M_B^2$ in Fig.~\ref{figsA} and \ref{figsD} for currents (\ref{Ja}) and (\ref{Jd}), respectively. Note that with the replacement of $q$-quark by $s$-quark, the currents in (\ref{Ja})-(\ref{Jc}) keep the same.

The errors in results (\ref{m1})-(\ref{m6}) mainly stem from the uncertainties in quark masses, condensates and threshold parameter $\sqrt{s_0}$. For the convenience of reference, a collection of continuum thresholds, Borel parameters, and predicted masses of $1^{-+}$ light tetraquark states are tabulated in Table \ref{mass}.

\begin{table}
\begin{center}
\renewcommand\tabcolsep{10pt}
\caption{The continuum thresholds, Borel parameters, and predicted masses of tetraquark states.}\label{mass}
\begin{tabular}{ccccc}\hline\hline
Configuration                                                      &Current   & $\sqrt{s_0}\;(\text{GeV})$     &$M_B^2\;(\text{GeV}^2)$ &$M^X\;(\text{GeV})$       \\ \hline
$[1_c]_{\bar{s}s}\otimes[1_c]_{\bar{q}q}$           &$A$        & $2.2\pm0.1$                             &$1.1-1.6$                      &$1.87\pm0.08$         \\
                                                                            &$B$        & $2.1\pm0.1$                             &$1.1-1.6$                      &$1.75\pm0.08$          \\\hline
$[1_c]_{\bar{s}q}\otimes[1_c]_{\bar{q}s}$           &$C$        & $2.4\pm0.1$                             &$1.3-1.8$                      &$2.05\pm0.07$           \\
                                                                            &$D$        & $2.1\pm0.1$                             &$1.2-1.7$                      &$1.63\pm0.12$          \\\hline
$[1_c]_{\bar{s}s}\otimes[1_c]_{\bar{s}s}$           &$A$         & $2.5\pm0.1$                            &$1.3-1.9$                        &$2.14\pm0.07$        \\
                                                                            &$D$        & $2.2\pm0.1$                             &$1.3-1.8$                       &$1.71\pm0.11$          \\
\hline
 \hline
\end{tabular}
\end{center}
\end{table}

\subsection{Decay analyses}\label{decay}

The preceding numerical results indicate that the $\eta_1(1855)$ is close in magnitude to our calculation of current in Eq.(\ref{Ja}) light tetraquark state (tetraquark $A$), which is to say $\eta_1(1855)$ at least has a large component in tetraquark configuration. To finally pin down the inner structure of $\eta_1(1855)$, a feasible procedure is to scrutinize its various decay products. Under the requirements of parity, angular momentum and isospin conservation, the $S$-wave $\eta f_1(1285)$, $P$-wave $\eta\eta^\prime$ and $\eta\eta(1295)$ decay processes may readily proceed for tetraquark $A$. There are at least two main differences between the decay modes of tetraquark $A$ and hybrid state:
\begin{enumerate}
\item  Decay modes of $\pi a_1(1260)$ and $\pi a_1(1640)$ for tetraquark $A$ are suppressed by the $s$-quarks translating to $q$-quarks, while the hybrid may decay to $\pi a_1$ more straightforwardly.
 \item  The tetraquark $A$ in configuration $[1_c]_{\bar{s}s}\otimes[1_c]_{\bar{q}q}$ is relatively tamed to decay to $K_1(1270)\bar{K}$, which is in configuration $[1_c]_{\bar{s}q}\otimes[1_c]_{\bar{s}q}$, while for a hybrid this decay channel looks more accessible.
\end{enumerate}

With above naive analyses, we expect future experimental measurement on $\eta_1(1855)$ decays via these channels may help to decode more of the $\eta_1(1855)$ inner structure.

\section{Summary}

In summary, we have investigated the light tetraquark states in the configurations of $[1_c]_{\bar{s}s}\otimes[1_c]_{\bar{q}q}$, $[1_c]_{\bar{s}q}\otimes[1_c]_{\bar{q}s}$, and $[1_c]_{\bar{s}s}\otimes[1_c]_{\bar{s}s}$ with $J^{PC}=1^{-+}$ in the framework of QCD sum rules. Our numerical results are tabulated in Table \ref{mass}. Results indicate that the $\eta_1(1855)$ reported by BESIII is close in magnitude to our calculation of current (\ref{Ja}) light tetraquark state. The possible tetraquark and hybrid decay modes are analyzed, through which the inner structure of $\eta_1(1855)$ may be elucidated with the experimental verification. Moreover, we also predict theoretically the possible existence of five other $1^{-+}$ tetraquark states.

Finally, it should be noted that the masses of hadronic states obtained through currents Eqs.~(\ref{Ja})-(\ref{Jd}) lie closely to each other, say semidegenerate, which poses a challenge to experiment on how to disentangle them.

\vspace{.5cm} {\bf Acknowledgments} \vspace{.5cm}

This work was supported in part by the National Key Research and Development Program of China under Contracts Nos. 2020YFA0406400, and the National Natural Science Foundation of China (NSFC) under the Grants 11975236 and 11635009.


\begin{widetext}
\appendix

\section{THE SPECTRAL DENSITIES FOR CASE $A$-$D$}

For the current in Eq. (\ref{Ja}), we obtain the spectral densities as follows:
\begin{eqnarray}
\rho^{pert}(s)&=&\frac{s^3(96m_s^2+5s)-120m_q^2s^2(12m_s^2+s)} {2^{15}\times3\times5\;\pi^6}\;,\\
\rho^{\langle \bar{q} q\rangle}(s)&=&\frac{(64m_s^2+11s)s m_q{\langle \bar{q} q\rangle}+3(8m_q^2-s)sm_s{\langle \bar{s} s\rangle}}{2^{10}\;\pi^4}\;,\\
\rho^{\langle g_s^2 G^2\rangle}(s)&=&-\frac{23s(18m_q^2+8m_s^2-s)\langle g_s^2 G^2\rangle}{2^{13}\times3^2\; \pi^6}\;,\\
\rho^{\langle g_s \bar{q} \sigma \cdot G q\rangle}(s)&=&-\frac{5m_q(54m_s^2+25s){\langle g_s \bar{q} \sigma \cdot G q\rangle}}{2^8\times3^2\;\pi^4}-\frac{25m_s(3m_q^2-s){\langle g_s\bar{s} \sigma \cdot G s\rangle}}{3\times2^7\;\pi^4}\;,\\
\rho^{\langle \bar{q} q\rangle^2}(s)&=&\frac{(m_q^2-8m_s^2-4s){\langle \bar{q} q\rangle^2}}{2^7\;\pi^2}+\frac{(8s+3m_s^2-24m_q^2){\langle \bar{s} s\rangle^2}}{3\times2^7\;\pi^2}\nonumber\\
&-&\frac{5m_qm_s{\langle \bar{q} q\rangle}{\langle \bar{s} s\rangle}}{2^5\;\pi^2}\;,\\
\rho^{\langle \bar{q} q\rangle \langle g_s^2 G^2\rangle}(s)&=&\frac{115m_q{\langle \bar{q} q\rangle\langle g_s^2 G^2\rangle}+23m_s{\langle \bar{s} s\rangle\langle g_s^2 G^2\rangle}}{3\times2^{10}\;\pi^2}\;,\\
\rho^{\langle \bar{q} q\rangle \langle g_s \bar{q} \sigma \cdot G q\rangle}(s)&=&\frac{2{\langle \bar{q} q\rangle\langle g_s \bar{q} \sigma \cdot G q\rangle}-{\langle \bar{s} s\rangle\langle g_s \bar{s} \sigma \cdot G s\rangle}}{2^6\;\pi^2}\;,\\
\Pi^{\langle \bar{q} q\rangle^2}(M_B^2)&=&-\frac{m_q^2m_s^2{\langle \bar{s} s\rangle^2}}{2^6\;\pi^2}\;,\\
\Pi^{\langle \bar{q} q\rangle \langle g_s \bar{q} \sigma \cdot G q\rangle}(M_B^2)&=&\frac{m_s^2{\langle \bar{q} q\rangle\langle g_s \bar{q} \sigma \cdot G q\rangle}}{2^5\;\pi^2}+\frac{55m_qm_s{\langle \bar{s} s\rangle\langle g_s \bar{q} \sigma \cdot G q\rangle}}{2^5\times3^2\;\pi^2}+\frac{25m_qm_s{\langle \bar{q} q\rangle\langle g_s \bar{s} \sigma \cdot G s\rangle}}{2\times2^4\;\pi^2}\nonumber\\
&+&\frac{m_q^2{\langle \bar{s} s\rangle\langle g_s \bar{s} \sigma \cdot G s\rangle}}{2^5\;\pi^2}+\frac{m_s^2m_q^2{\langle \bar{q} q\rangle\langle g_s \bar{q} \sigma \cdot G q\rangle}}{3\times2^6\;\pi^2\;M_B^2}\;.
\end{eqnarray}

For the current in Eq. (\ref{Jb}), the spectral densities will be the same with the current in Eq. (\ref{Ja}) but with letters $q$ and $s$ to be exchanged.

For the current in Eq. (\ref{Jc}), we obtain the spectral densities as follows:
\begin{eqnarray}
\rho^{pert}(s)&=&\frac{s^2 \left(-12 m_q^2 \left(120 m_s^2+s\right)-48 s m_q m_s+s \left(5 s-12 m_s^2\right)\right)}{2^{15}\times3\times5\; \pi ^6}\;,\\
\rho^{\langle \bar{q} q\rangle}(s)&=&\frac{\langle \bar{q} q\rangle s \left(-4 m_q^2 m_s+2 m_q \left(11 m_s^2+s\right)+s m_s\right)}{2^9\; \pi ^4}+ \nonumber\\
&&\frac{\langle \bar{s} s\rangle s \left(22 m_q^2 m_s+m_q \left(s-4 m_s^2\right)+2 s m_s\right)}{2^9\; \pi ^4}\;,\\
\rho^{\langle g_s^2 G^2\rangle}(s)&=&\frac{23 \langle g_s^2 G^2\rangle s \left(s-26 m_q m_s\right)}{2^{13}\times 3^2\; \pi ^6}   \;,\\
\rho^{\langle g_s \bar{q} \sigma \cdot G q\rangle}(s)&=&\frac{\langle g_s \bar{q} \sigma \cdot G q\rangle \left(-603 m_q^2 m_s-99 m_q m_s^2+94 s m_q+51 s m_s\right)}{2^9\times 3^2\; \pi ^4}+ \nonumber\\
&&\frac{\langle g_s \bar{s} \sigma \cdot G s\rangle \left(-99 m_q^2 m_s-603 m_q m_s^2+51 s m_q+94 s m_s\right)}{2^9\times 3^2\; \pi ^4}   \;,\\
\rho^{\langle \bar{q} q\rangle^2}(s)&=&\frac{\langle \bar{q} q\rangle^2 \left(12 m_q m_s-2 \left(12 m_s^2+s\right)+3 m_q^2\right)}{2^7\times 3\; \pi ^2}+ \nonumber\\
&&\frac{\langle \bar{q} q\rangle \langle \bar{s} s\rangle \left(-15 m_q m_s+3 m_q^2+3 m_s^2-2\right)}{2^5\times 3\; \pi ^2}- \nonumber\\
&&\frac{\langle \bar{s} s\rangle^2 \left(-12 m_q m_s+24 m_q^2-3 m_s^2+2 s\right)}{2^7\times 3\; \pi ^2}  \;,\\
\rho^{\langle \bar{q} q\rangle \langle g_s^2 G^2\rangle}(s)&=&\frac{23 \langle \bar{q} q\rangle \langle g_s^2 G^2\rangle m_s}{2^{10}\; \pi ^4}+\frac{23 \langle \bar{s} s\rangle \langle g_s^2 G^2\rangle m_q}{2^{10}\; \pi ^4}   \;,\\
\rho^{\langle \bar{q} q\rangle \langle g_s \bar{q} \sigma \cdot G q\rangle}(s)&=&\frac{\langle \bar{q} q\rangle \langle g_s \bar{q} \sigma \cdot G q\rangle \left(299 m_q m_s-46 m_q^2+9\right)}{2^7\times 3^2\; \pi ^2}+\nonumber\\
&&\frac{\langle \bar{q} q\rangle \langle g_s \bar{s} \sigma \cdot G s\rangle \left(23 m_q m_s+230 m_s^2+18\right)}{2^7\times 3^2\; \pi ^2}+\nonumber\\
&&\frac{\langle \bar{s} s\rangle \langle g_s \bar{q} \sigma \cdot G q\rangle \left(23 m_q m_s+230 m_q^2+18\right)}{2^7\times 3^2\; \pi ^2}+\nonumber\\
&&\frac{\langle \bar{s} s\rangle \langle g_s \bar{s} \sigma \cdot G s\rangle \left(299 m_q m_s-46 m_s^2+9\right)}{2^7\times 3^2\; \pi ^2}  \;,\\
\Pi^{\langle \bar{q} q\rangle^2}(M_B^2)&=&-\frac{\langle \bar{q} q\rangle^2 m_q^2 m_s^2}{2^{7}\; \pi ^2}-\frac{\langle \bar{q} q\rangle \langle \bar{s} s\rangle m_q^2 m_s^2}{2^{5}\; \pi ^2}-\frac{\langle \bar{s} s\rangle^2 m_q^2 m_s^2}{2^{7}\; \pi ^2}   \;,\\
\Pi^{\langle \bar{q} q\rangle \langle g_s \bar{q} \sigma \cdot G q\rangle}(M_B^2)&=&\frac{\langle \bar{q} q\rangle \langle g_s \bar{q} \sigma \cdot G q\rangle \left(269 M_B^2 m_q m_s-46 M_B^2 m_q^2+36 M_B^2 m_s^2+3 m_q^2 m_s^2\right)}{2^7\times 3^2\; \pi ^2 M_B^2}+\nonumber\\
&&\frac{\langle \bar{q} q\rangle \langle g_s \bar{s} \sigma \cdot G s\rangle \left(65 M_B^2 m_q m_s-18 M_B^2 m_q^2+218 M_B^2 m_s^2+6 m_q^2 m_s^2\right)}{2^7\times 3^2\; \pi ^2 M_B^2}+\nonumber\\
&&\frac{\langle \bar{s} s\rangle \langle g_s \bar{q} \sigma \cdot G q\rangle \left(65 M_B^2 m_q m_s+218 M_B^2 m_q^2-18 M_B^2 m_s^2+6 m_q^2 m_s^2\right)}{2^7\times 3^2\; \pi ^2 M_B^2}+\nonumber\\
&&\frac{\langle \bar{s} s\rangle \langle g_s \bar{s} \sigma \cdot G s\rangle \left(269 M_B^2 m_q m_s+36 M_B^2 m_q^2-46 M_B^2 m_s^2+3 m_q^2 m_s^2\right)}{2^7\times 3^2\; \pi ^2 M_B^2}  \nonumber \;.\\
\end{eqnarray}

For the current in Eq. (\ref{Jd}), we obtain the spectral densities as follows:
\begin{eqnarray}
	\rho^{pert}(s)&=&\frac{s^2 \left(-12 m_q^2 \left(120 m_s^2+s\right)+48 s m_q m_s+s \left(5 s-12 m_s^2\right)\right)}{2^{15}\times3\times5\; \pi ^6}\;,\\
	\rho^{\langle \bar{q} q\rangle}(s)&=&\frac{\langle \bar{s} s\rangle s \left(22 m_q^2 m_s+4 m_q m_s^2-s m_q+2 s m_s\right)}{2^9\; \pi ^4}+\nonumber\\
	&&\frac{\langle \bar{q} q\rangle s \left(4 m_q^2 m_s+2 m_q \left(11 m_s^2+s\right)-s m_s\right)}{2^9\; \pi ^4}\;,\\
	\rho^{\langle g_s^2 G^2\rangle}(s)&=&\frac{23 \langle g_s^2 G^2\rangle s \left(26 m_q m_s+s\right)}{2^{13}\times 3^2\; \pi ^6}   \;,\\
	\rho^{\langle g_s \bar{q} \sigma \cdot G q\rangle}(s)&=&\frac{\langle g_s \bar{q} \sigma \cdot G q\rangle \left(603 m_q^2 m_s-99 m_q m_s^2+94 s m_q-51 s m_s\right)}{2^9\times 3^2\; \pi ^4}+\nonumber\\
	&&\frac{\langle g_s \bar{s} \sigma \cdot G s\rangle \left(-99 m_q^2 m_s+603 m_q m_s^2-51 s m_q+94 s m_s\right)}{2^9\times 3^2\; \pi ^4}  \;,\\
	\rho^{\langle \bar{q} q\rangle^2}(s)&=&\frac{\langle \bar{q} q\rangle^2 \left(-12 m_q m_s+3 m_q^2-2 \left(12 m_s^2+s\right)\right)}{2^7\times 3\; \pi ^2}+\nonumber\\
	&&\frac{\langle \bar{q} q\rangle \langle \bar{s} s\rangle \left(-15 m_q m_s-3 m_q^2-3 m_s^2+2\right)}{2^5\times 3\; \pi ^2}-\nonumber\\
	&&\frac{\langle \bar{s} s\rangle^2 \left(12 m_q m_s+24 m_q^2-3 m_s^2+2 s\right)}{2^7\times 3\; \pi ^2}  \;,\\
	\rho^{\langle \bar{q} q\rangle \langle g_s^2 G^2\rangle}(s)&=&-\frac{23 \langle \bar{q} q\rangle \langle g_s^2 G^2\rangle m_s}{2^{10}\; \pi ^4}-\frac{23 \langle \bar{s} s\rangle \langle g_s^2 G^2\rangle m_q}{2^{10}\; \pi ^4}   \;,\\
	\rho^{\langle \bar{q} q\rangle \langle g_s \bar{q} \sigma \cdot G q\rangle}(s)&=&\frac{-\langle \bar{q} q\rangle \langle g_s \bar{q} \sigma \cdot G q\rangle \left(299 m_q m_s+46 m_q^2-9\right)}{2^7\times 3^2\; \pi ^2}-\nonumber\\
	&&\frac{\langle \bar{q} q\rangle \langle g_s \bar{s} \sigma \cdot G s\rangle \left(-23 m_q m_s+230 m_s^2+18\right)}{2^7\times 3^2\; \pi ^2}-\nonumber\\
	&&\frac{\langle \bar{s} s\rangle \langle g_s \bar{q} \sigma \cdot G q\rangle \left(-23 m_q m_s+230 m_q^2+18\right)}{2^7\times 3^2\; \pi ^2}-\nonumber\\
	&&\frac{\langle \bar{s} s\rangle \langle g_s \bar{s} \sigma \cdot G s\rangle \left(299 m_q m_s+46 m_s^2-9\right)}{2^7\times 3^2\; \pi ^2}  \;,\\
	\Pi^{\langle \bar{q} q\rangle^2}(M_B^2)&=&-\frac{\langle \bar{q} q\rangle^2 m_q^2 m_s^2}{2^{7}\; \pi ^2}+\frac{\langle \bar{q} q\rangle \langle \bar{s} s\rangle m_q^2 m_s^2}{2^{5}\; \pi ^2}-\frac{\langle \bar{s} s\rangle^2 m_q^2 m_s^2}{2^{7}\; \pi ^2}   \;,\\
	\Pi^{\langle \bar{q} q\rangle \langle g_s \bar{q} \sigma \cdot G q\rangle}(M_B^2)&=&\frac{\langle \bar{q} q\rangle \langle g_s \bar{q} \sigma \cdot G q\rangle \left(m_q^2 \left(\frac{3 m_s^2}{M_B^2}-46\right)-269 m_q m_s+36 m_s^2\right)}{2^7\times 3^2\; \pi ^2}+\nonumber\\
	&&\frac{\langle \bar{q} q\rangle \langle g_s \bar{s} \sigma \cdot G s\rangle \left(-6 m_q^2 \left(\frac{m_s^2}{M_B^2}-3\right)+65 m_q m_s-218 m_s^2\right)}{2^7\times 3^2\; \pi ^2}+\nonumber\\
	&&\frac{\langle \bar{s} s\rangle \langle g_s \bar{q} \sigma \cdot G q\rangle \left(-2 m_q^2 \left(\frac{3 m_s^2}{M_B^2}+109\right)+65 m_q m_s+18 m_s^2\right)}{2^7\times 3^2\; \pi ^2}+\nonumber\\
	&&\frac{\langle \bar{s} s\rangle \langle g_s \bar{s} \sigma \cdot G s\rangle \left(3 m_q^2 \left(\frac{m_s^2}{M_B^2}+12\right)-269 m_q m_s-46 m_s^2\right)}{2^7\times 3^2\; \pi ^2}   \;.
\end{eqnarray}

\end{widetext}
\end{document}